\documentclass{jpsj2}
\title
{Perturbation Theory on the Superconductivity of Heavy Fermion
Superconductors ${\rm CeIr}_{x}{\rm Co}_{1-x}{\rm In}_{5}$}

\author
{ 
Yunori  {\sc Nisikawa}\footnote{E-mail
address:nisikawa@ton.scphys.kyoto-u.ac.jp}, Hiroaki {\sc Ikeda}
and Kosaku {\sc Yamada}
}

\inst{Department of Physics, Kyoto University, Kyoto 606-8502}

\abst{We reformulate the \'Eliashberg's equation 
for the superconducting transition within the quasi-particle
description, which takes account of the heavy electron mass.
We discuss the superconductivity 
of ${\rm CeIr}_{x}{\rm Co}_{1-x}{\rm In}_{5}$ 
by using such a renormalized formula.
On the basis of an effective two-dimensional Hubbard model 
which represents the most heavy quasi-two dimensional
Fermi surface with $f$-character 
of ${\rm CeIr}_{x}{\rm Co}_{1-x}{\rm In}_{5}$,
both normal and anomalous self-energies are calculated up to 
third order with respect to the renormalized on-site repulsion $U$ 
between quasi-particles. The superconducting
transition temperature is obtained by solving the
${\rm\acute{E}}$liashberg's equation.
Reasonable transition temperatures are obtained 
for moderately large $U$.
It is found that the momentum and frequency dependence of 
spin fluctuations given by RPA-like terms gives rise to the d-wave
pairing state, while 
the vertex correction terms are important for obtaining reasonable 
transition temperatures and 
for differentiating $n$-dependence of $T_{{\rm c}}$.
These obtained results seem to explain the $x$-dependence of $T_{{\rm c}}$
in ${\rm CeIr}_{x}{\rm Co}_{1-x}{\rm In}_{5}$ and confirm the
usefulness of the perturbation theory with respect to $U$.}

\kword{heavy fermion systems, electron correlation, 
${\rm Ce(Ir,Co,Rh)In_{5}}$, superconducting transition temperature, 
perturbation approach, vertex correction}

\begin{document}
\maketitle

\section{Introduction}
In heavy fermion system,
the effective mass is enhanced by $10^{2}\sim 10^{3}$ times
as large as that of free electron.
The coefficient of $T^{2}$-law of resistivity
is also enhanced by $10^{4}\sim 10^{6}$ times as large as that of 
conventional  metal.
Such a large electron mass
and coefficient of $T^{2}$-law of resistivity
stem from the electron correlation.
In heavy fermion superconductors,
itinerant electrons compose such a Fermi liquid 
and then undergo the superconducting transition.
From this point of view, we can expect that superconductivities in heavy fermion
systems are derived also from the electron correlation
through the momentum and frequency dependence of the effective
interaction between electrons.
Therefore, in contrast to the cuprates,
it is not clear whether taking only
the effect of the specific diagrams which are favorable
to the spin fluctuations is reliable
for revealing the mechanism of 
superconductivity in the heavy fermion superconductors.

In this paper, we present an essentially important point in theoretical study of
superconductivity in heavy fermion systems.
In contrast to the cuprates, the momentum dependence of the mass
enhancement factor in heavy fermion systems is weak because
of the localized nature of $f$-electrons.
In this case we can reformulate the ${\rm\acute{E}}$liashberg's
 equation also in the
renormalized form, where we have already included the large mass
enhancement effect.
By using the renormalized Fermi energy as unit of energy 
the critical temperature $T_{{\rm c}}$ determined by the
perturbation is the same as that obtained after taking
account of the mass enhancement effect.
Thus, we can formulate the perturbation theory which is consistent with 
the large mass enhancement.
The formulation is one of purposes of the present paper.
Therefore, we calculate the superconducting 
transition temperature $T_{{\rm c}}$ of heavy fermion 
superconductors by the perturbation theory based on Fermi liquid theory.
In our formalism, the effective interaction corresponding to 
the superexchange interaction
~\cite{rf:Ohkawa}
 is included in the
vertex terms in principle,
 because we start from renormalized Fermi liquid and treat 
only the Coulomb repulsion by the perturbation theory.

The perturbation approach is sensitive to the
dispersion of the bare energy band by its nature,
it implies that the lattice structures and the band
filling play the essential roles in the calculation of $T_{{\rm c}}$.
Therefore it is important to evaluate $T_{{\rm c}}$ 
on the basis of the detailed electronic structure in each system.
In this paper, we calculate $T_{{\rm c}}$ 
of ${\rm CeIr}_{x}{\rm Co}_{1-x}{\rm In}_{5}$.

The organization of this paper is as follows. 
In $\S 2$, a theoretical treatment of heavy fermion
superconductivity is presented.
In $\S 3$, $T_{{\rm c}}$ of ${\rm CeIr}_{x}{\rm Co}_{1-x}{\rm In}_{5}$
 is calculated, after giving the  model Hamiltonian, 
perturbation expansion terms and ${\rm \acute{E}}$liashberg's equation. 
Moreover the selfenergy, density of states 
and other quantities are shown. 
Finally in $\S 4$, summary, discussions and conclusion are presented.
\section{A theory of heavy fermion superconductivity}
In this section, we discuss a theoretical treatment of heavy fermion
superconductivity.
As indicated in various experiments, such as a jump of the specific heat
at the transition temperature ($T_{\rm c}$), the superconductivity can be
interpreted as a transition of quasi-particles with heavy electron mass.
We here formulate the way to calculate the superconducting transition
temperature $T_{\rm c}$ on the heavy fermion quasi-particle state
in the periodic Anderson model (PAM).

One of the characteristics in heavy fermion systems is the weak momentum
dependence, corresponding to degree of the local 
$f$-electron spins at high
temperatures or energies.
The perturbation expansion in the hybridization matrix element in PAM leads to
the RKKY interaction between these local spins.
This expansion is valid at high temperatures, and may be the origin of the
magnetic transition in $f$-electron systems.
However, if no magnetic transition occurs, the system goes to a singlet
ground state as the whole.
This just is the Fermi liquid state with heavy electron mass.
In this case, we had better treat it by the perturbation theory with respect to
the on-site Coulomb repulsion $U$ between $f$-electrons,
because of the analyticity about $U$.
The analytic property has been exactly proved in impurity Anderson model~\cite{rf:exact1,rf:exact2,rf:exact3,rf:ana}.
Although there is no exact such proof in PAM, the principle of the
adiabatic continuation confirms the correctness of
the perturbation theory in $U$.

 From such a point of view, one of the authors (K.Y.) has described
the Fermi liquid theory for the heavy fermion state in PAM~\cite{rf:K-W}.
The momentum independent part of the 4-point vertex functions
$\Gamma_{\sigma\sigma'}^{\rm loc}$ between 
$f$-electrons with the spin $\sigma$
and $\sigma'$ plays an important role.
This s-wave scattering part, corresponding to the above-mentioned
local character at high energies, leads to the featureless large mass
enhancement factor $\tilde\gamma=a^{-1}$.
In fact, when the momentum dependence of the mass enhancement factor
can be ignored, the $T$-linear coefficient of the specific heat in PAM,
is given by
$\gamma \simeq (2/3)\pi^2k_{\rm B}^2{\rho^f(0)}^2
\Gamma_{\uparrow\downarrow}^{\rm loc}
= (2/3)\pi^2k_{\rm B}^2\rho^f(0)\tilde\gamma$, where
$k_{\rm B}$ is the Boltzmann's constant, and $\rho^f(0)$ the $f$-electron
density of states at the Fermi level.
This indicates that the large mass enhancement factor $\tilde\gamma=a^{-1}$
is represented as $\rho^f(0)\Gamma_{\uparrow\downarrow}^{\rm loc}$
with use of the s-wave scattering part of the 4-point vertex.
Thus, we can see that the vertex $\Gamma_{\uparrow\downarrow}^{\rm loc}$ is
enhanced by $\tilde\gamma=a^{-1}$.
This corresponds to the fact that the interaction
$a^2\Gamma_{\uparrow\downarrow}^{\rm loc}$ between the quasi-particles has
the order of magnitude of the effective band-width of quasi-particles
$T_0 \simeq 1/\tilde\gamma\rho^f(0)$;
$T_0$ is the characteristic energy scale of the heavy fermion state, and
below $T_0$, the low energy excitation can be described by the Fermi
liquid theory.
On the other hand, the imaginary part of the $f$-electron self-energy,
which is proportional to the $T^2$-term of the electrical resistivity,
is given by $\Delta_k=(4\pi^2/3)\rho^f(0)^2\pi
{\Gamma_{\uparrow\downarrow}^{\rm loc}}^{2}T^{2}$,
if the momentum dependence of the vertices can be ignored.
Thus, the coefficient $A$ of the $T^2$-term of the electrical resistivity is
proportional to $\gamma^2$ through the vertex
$\Gamma_{\uparrow\downarrow}^{\rm loc}$.
This relation $A \propto \gamma^2$ is just the Kadowaki-Wood's relation,
which holds in the case where the momentum dependence of the vertices is
sufficiently weak to be ignored.
Actually, in the heavy fermion systems, the Kadowaki-Wood's relation is
confirmed.
This fact means that the 4-point vertex function, i.e., the interaction 
between quasi-particles possesses the above-mentioned local nature.

The superconductivity in the heavy fermion systems appears under this
situation below $T_0$.
The s-wave large repulsive part $\Gamma_{\uparrow\downarrow}^{\rm loc}$
prevents the appearance of the s-wave singlet superconductivity
in strongly correlated systems.
In this case, an anisotropic (such as p-, d-wave and so on) superconductivity will
be realized due to the remaining momentum dependence of the 4-point vertices.
We here discuss such a transition to the anisotropic superconductivity
on the basis of the 
heavy fermion quasi-particle state renormalized by the large s-wave
scattering part $\Gamma_{\uparrow\downarrow}^{\rm loc}$.

First of all, we divide the 4-point full vertex function
$\Gamma_{\sigma\sigma'}(p_1,p_2;p_3,p_4)$
into the large s-wave scattering part $\Gamma_{\sigma\sigma'}^{\rm loc}$
and the non-s wave part $\Delta\Gamma_{\sigma\sigma'}(p_1,p_2;p_3,p_4)$;
\begin{equation}
\Gamma_{\sigma\sigma'}(p_1,p_2;p_3,p_4)=\Gamma_{\sigma\sigma'}^{\rm loc}
+\Delta\Gamma_{\sigma\sigma'}(p_1,p_2;p_3,p_4),
\end{equation}
where $p_1$ and $p_2$ are the incident momenta, and $p_3$, $p_4$ the outgoing.
The momentum dependent part has rather remarkable momentum dependence
as the heavy fermion quasi-particles grow below $T_0$.
The effective Cooper pairing interaction in the anisotropic superconductivity
results from the momentum dependence of
$\Delta\Gamma_{\sigma\sigma'}(p_1,p_2;p_3,p_4)$.
Our purpose is to formulate how to calculate
$\Delta\Gamma_{\sigma\sigma'}(p_1,p_2;p_3,p_4)$
for the heavy fermion quasi-particles, which are renormalized
by the large local part $\Gamma_{\uparrow\downarrow}^{\rm loc}$.
Corresponding to these two terms of the full vertex function,
we can also separate the self-energy into the local part and the non-local part
\begin{equation}
\Sigma(k,\omega)=\Sigma_{\rm loc}(\omega)+\Delta\Sigma(k,\omega).
\end{equation}
In this case, the $f$-electron Green's function below $T_0$ is given by
\begin{eqnarray}
\hspace{-8mm} G(k,\omega)&=&
\frac{1}{\omega-\xi_k-\Sigma^{\rm loc}(\omega)-\Delta\Sigma(k,\omega)
-\frac{V_k^2}{\omega-\epsilon_k}},
\nonumber \\
&=&\frac{a}{\omega-\tilde{E}_k-\tilde{\Sigma}(k,\omega)
-\frac{\tilde{V}_k^2}{\omega-\epsilon_k}}+G_{\rm inc}(\omega),
\end{eqnarray}
where $a$ is a wave function renormalization factor, and
$\tilde{E}_k=a(\xi_k+{\rm Re}\Sigma^{\rm loc}(0))$, $\tilde{V}_k=\sqrt{a}V_k$
and $\tilde{\Sigma}(k,\omega)=a\Delta\Sigma(k,\omega)$ are, respectively,
a renormalized $f$-electron band, hybridization term and self-energy.
$G_{\rm inc}(\omega)$ is a featureless incoherent part.
The imaginary part 
of $\Sigma^{\rm loc}$ proportional to $\omega^2$ can be
included in the remaining renormalized self-energy $\tilde{\Sigma}(k,\omega)$,
as indicated by Hewson for the impurity Anderson model~\cite{rf:Hew1,rf:Hew2}.
 From the Ward-Takahashi identity, we can obtain the large mass enhancement
factor,
\begin{equation}
\tilde{\gamma}=a^{-1}=1-\dfrac{d\Sigma_{\rm loc}(0)}{d\omega}
=1-\dfrac{i}{2}\int d\omega\Gamma_{\uparrow\downarrow}^{\rm loc}G_{\rm inc}^2(\omega).
\end{equation}
The above equation of the Green's function is the counterpart
on the quasi-particle description.
In addition, we can set the large local vertex
$\Gamma_{\uparrow\downarrow}^{\rm loc}$ as a constant value at zero
frequencies, as far as the frequency of external lines is less than $T_0$.
In this case, the effective on-site interaction
$a^2\Gamma_{\uparrow\downarrow}^{\rm loc}$ works between the quasi-particles.
In the impurity Anderson model,
$a^2\Gamma_{\uparrow\downarrow}^{\rm loc}=a\pi\Delta=4T_{\rm K}$,
where $\Delta$ and $T_{\rm K}$ are, respectively, the width of the virtual
bound state and the Kondo temperature.
Therefore, we can assume that $a^2\Gamma_{\uparrow\downarrow}^{\rm loc}$
has the order of $T_0$.
This is also consistent with the above-mentioned discussion of the coefficient
of the specific heat.

\begin{figure}
\includegraphics[width=0.6 \linewidth]{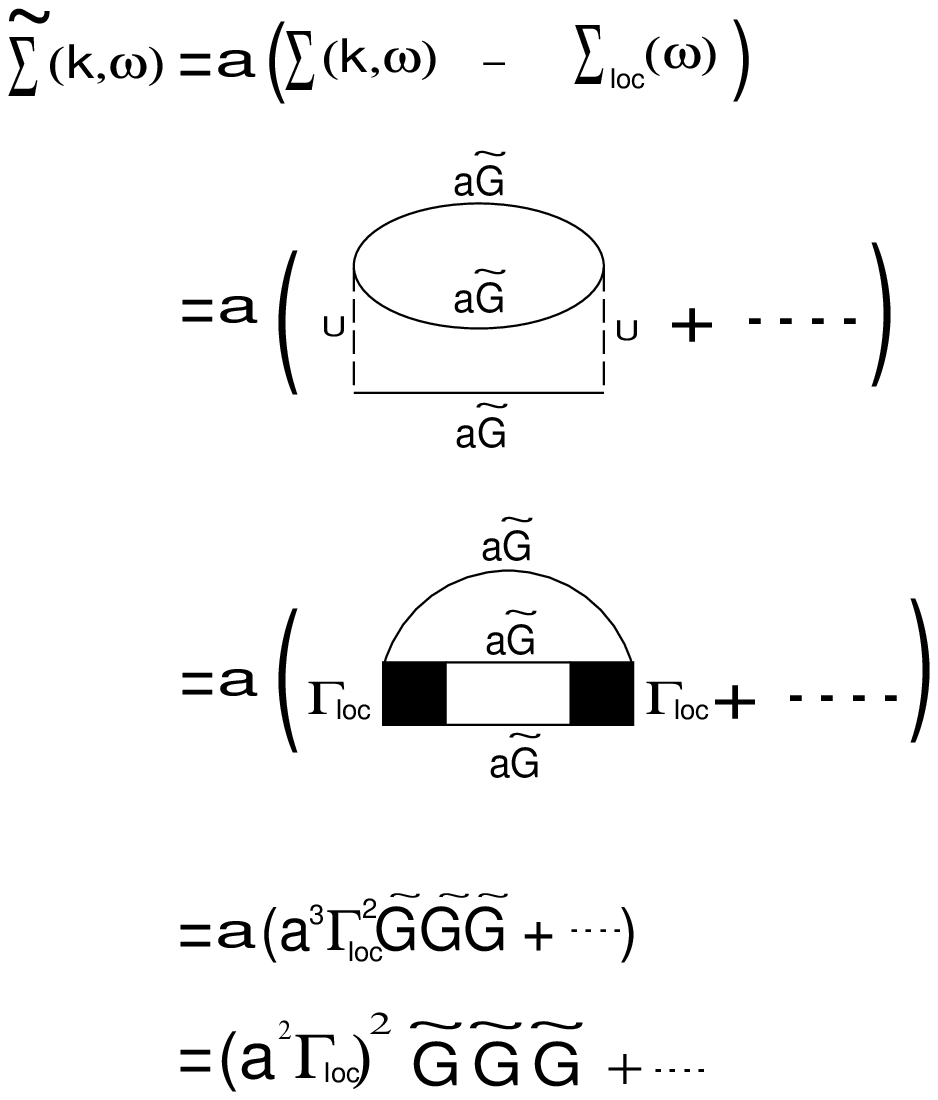}
\caption{The renormalized self-energy $\tilde{\Sigma}({\bf k},\omega)$
 can be re-written as the expansion with respect to 
the quasi-particle Green's function $\tilde{G}$ 
and the quasi-particle s-wave scattering interaction
$\tilde{\Gamma}_{\rm loc}=a^{2}\Gamma_{{\rm loc}}$}
\label{fig:kurikomi1}
\end{figure}

\begin{figure}
\includegraphics[width=0.8 \linewidth]{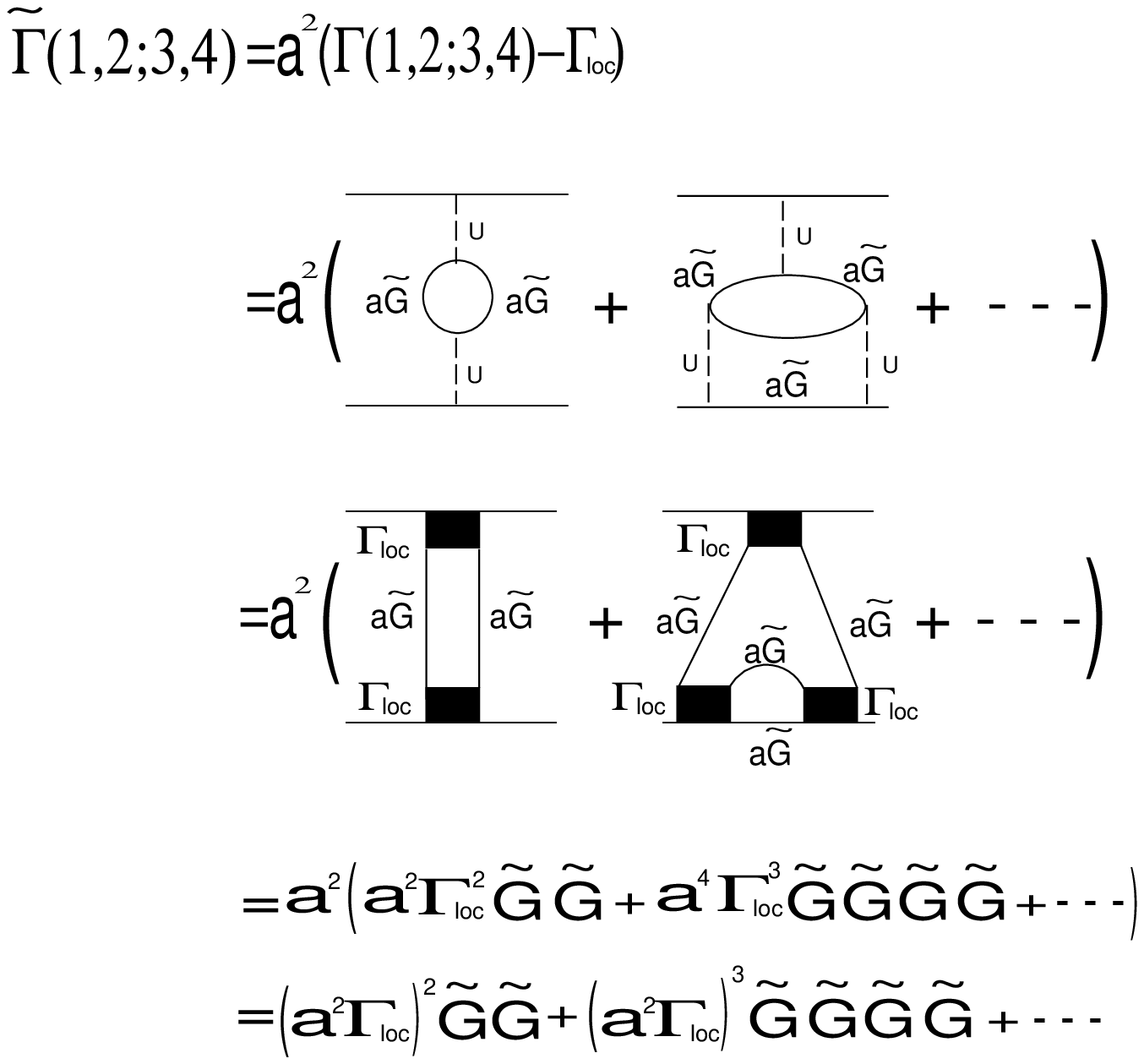}
\caption{The full vertex
$\tilde{\Gamma}(1,2;3,4)$ can be re-written
as the expansion with respect to the quasi-particle Green's function
$\tilde{G}$ and the quasi-particle s-wave scattering interaction
$\tilde{\Gamma}_{\rm loc}=a^{2}\Gamma_{{\rm loc}}$ }
\label{fig:kurikomi2}
\end{figure}

We now show that in the region $\omega \le T_0$,
the renormalized self-energy $\tilde{\Sigma}(k,\omega)$
and vertex function $\tilde{\Gamma}_{\sigma\sigma'}(p_1,p_2;p_3,p_4)=
a^2\Delta\Gamma_{\sigma\sigma'}(p_1,p_2;p_3,p_4)$
can be discussed with the perturbation expansion with respect to the
effective on-site interaction 
$\tilde{\Gamma}_{\uparrow\downarrow}^{\rm loc}=
a^2\Gamma_{\uparrow\downarrow}^{\rm loc}$.
This expansion can be also considered as the one
with respect to the momentum dependence which is governed by the
renormalized Fermi surface.
First, let us consider the renormalized self-energy
$\tilde{\Sigma}(k,\omega)=a\Delta\Sigma(k,\omega)$.
Because we have divided the Green's function into the two parts
$a\tilde{G}(k,\omega)+G_{\rm inc}(\omega)$,
we can correspondingly divide any diagrams in the expansion with respect 
to the bare interaction $U$ into diagrams with and without the 
momentum dependence.
The latter is contained in the local self-energy $\Sigma_{\rm loc}(\omega)$,
on the other hand, the former can be rewritten as the expansion
with respect to the effective interaction
$\tilde{\Gamma}_{\uparrow\downarrow}^{\rm loc}$
by concentrating diagrams with the same momentum dependence.
As shown in Fig.~\ref{fig:kurikomi1}, 
for instance, the $U^2$-type diagram has the same
momentum dependence as the 
${\Gamma_{\uparrow\downarrow}^{\rm loc}}^2$-type,
each vertex of which includes all the s-wave scattering process.
Because three lines of the renormalized Green's function $a\tilde{G}(k,\omega)$
yields $a^3$, and the renormalized self-energy $\tilde{\Sigma}(k,\omega)$
includes $a$, the ${\Gamma_{\uparrow\downarrow}^{\rm loc}}^2$-type
is estimated by calculating
$a^4{\Gamma_{\uparrow\downarrow}^{\rm loc}}^2\tilde{G}\tilde{G}\tilde{G}$.
This just corresponds to the diagram of the order two with respect to
the effective interaction $\tilde{\Gamma}_{\uparrow\downarrow}^{\rm loc}=
a^2\Gamma_{\uparrow\downarrow}^{\rm loc}$.
In the same way, the full vertex
$\tilde{\Gamma}_{\sigma\sigma'}(p_1,p_2;p_3,p_4)$ can be also re-written
as the expansion with respect to the quasi-particle Green's function
$\tilde{G}(k,\omega)$ and the quasi-particle s-wave scattering interaction
$\tilde{\Gamma}_{\sigma\sigma'}^{\rm loc}$ as shown in Fig.~\ref{fig:kurikomi2}.
Thus, we can discuss the phenomena below $T_0$ within the renormalized
effective PAM, so far as the frequency of external lines is less than $T_0$.
The validity of such a perturbation expansion with respect to the
renormalized interaction is shown by the fact that the fixed point
Hamiltonian in the renormalization group can be written
as the renormalized PAM.
In this case, the locality of the relevant interaction is very important.
In the heavy fermion systems, this condition almost holds
as shown in the Kadowaki-Wood's relation.
We formulate below the superconducting transition on the quasi-particle
description discussed above.

The superconducting transition is always marked by divergence of
the full vertex $\Gamma_{\sigma\sigma'}(p,-p;p',-p')\equiv
\Gamma_{\sigma\sigma'}(p,p')$.
We can rewrite it as divergence of the quasi-particle full vertex
$\tilde{\Gamma}_{\sigma\sigma'}(p,p')$.
As usual, we estimate the \'Eliashberg equation,
\begin{equation}\label{kuriEli}
\sum_{p'}\Gamma^{(2)}(p,p')|G(p',i\omega'_n)|^2
\Delta(p',i\omega'_n)=\Delta(p,i\omega_n),
\end{equation}
where $\Delta(p,i\omega_n)$ is an anomalous self-energy with a fermion
Matsubara frequency $\omega_n=(2n+1)\pi T$, and the Cooper pairing effective
interaction $\Gamma^{(2)}(p,p')$ is the particle-particle irreducible
vertex.
The Eq.(~\ref{kuriEli} ) includes the integral of $|G(p',i\omega'_n)|^2$.
The most important part of this integral comes from the part mediated
by quasi-particles $a^2|\tilde{G}(p',i\omega'_n)|^{2}$.
The \'Eliashberg equation can be re-written as
\begin{equation}\label{qp-kuriEli}
\sum_{p'}a^2\Gamma^{(2)}(p,p')|\tilde{G}(p',i\omega'_n)|^2
\Delta(p',i\omega'_n)=\Delta(p,i\omega_n).
\end{equation}
This $a^2\Gamma^{(2)}(p,p')$ is the particle-particle irreducible
vertex between quasi-particles with the external frequencies less than $T_0$.
As discussed above, for such a vertex, we can also apply the perturbation
expansion with respect to the renormalized on-site repulsion
$a^2\Gamma_{\uparrow\downarrow}^{\rm loc}$ between quasi-particles.

In order to treat the heavy fermion superconductivity, 
we have approximately introduced 
the quasi-particle state renormalized by the s-wave scattering part,
which itself does not yield the pairing interaction. 
Starting from the quasi-particle state, we calculate 
the momentum dependent 
interaction between quasi-particles by the perturbation theory and 
derive the anisotropic superconductivity.
It should be noted that even 
if the effective interaction depends on momentum as seen in the cuprates,
we can treat such a separation of the vertices in the momentum space,
as far as, at the final 
stage, everything is included in a consistent way.

\section{Calculating $T_{{\rm c}}$ of ${\rm CeIr}_{x}{\rm Co}_{1-x}{\rm In}_{5}$}
\subsection{Introduction to ${\rm Ce(Rh,Ir,Co)In_{5}}$}
Superconductivities are discovered in
a series of ${\rm CeMIn_{5}}$ tetragonal compounds for M=${\rm Co, 
Rh}$ or ${\rm Ir}$.
${\rm CeIrIn_{5}}$~\cite{rf:CeIr} and ${\rm CeCoIn_{5}}$~\cite{rf:CeCo}
are superconductors at ambient pressure.
The transition temperature $T_{{\rm c}}$ (and the electronic specific
heat coefficients $\gamma$) are 0.4K (680${\rm mJ/K^{2}mol}$),
and 2.3K (300-1000${\rm mJ/K^{2}mol}$), respectively.
In contrast, ${\rm CeRhIn_{5}}$~\cite{rf:PcCeRh}
 orders antiferromagnetically below
$T_{{\rm N}}=3.8$K, whereas the superconductivity is observed under
pressure, $p\ge 1.6$GPa. 
The observations of power law temperature dependence of
the specific heat, thermal conductivity, and NMR relaxation rate
have identified ${\rm Ce(Rh,Ir,Co)In_{5}}$ 
as unconventional superconductors with line nodes.
The thermal conductivity measurement 
in a magnetic field rotating within the 2D planes
also reveals that the superconducting gap symmetry of ${\rm CeCoIn_{5}}$ 
most likely belongs to $d_{x^{2}-y^{2}}$~\cite{rf:dx2-y2}.
The temperature-composition phase diagram
for ${\rm Ce(Rh,Ir,Co)In_{5}}$ has been 
obtained mainly from heat capacity
measurements on single crystals~\cite{rf:phaseDia}. 
The phase diagram shows an interesting asymmetry with respect to
the coupling between magnetism and superconductivity: in the case of ${\rm
Rh}$-${\rm Ir}$, $T_{{\rm c}}$ increases as the magnetic boundary
 is approached, whereas in ${\rm Rh}$-${\rm Co}$ case, ${\rm T_{{\rm
 c}}}$ decreases.
According to the phase diagram, 
the superconducting phase of ${\rm CeIr}_{x}{\rm Co}_{1-x}{\rm In}_{5}$
 is apart from 
the antiferromagnetic ordered phase.
When $x$ increases from $0$ to $1$, $T_{{\rm c}}$ decreases
 from 2.3 K to 0.4 K.
The temperature dependence of the resistivity $\rho$ of  ${\rm CeIrIn_{5}}$
follows the dependence $\rho=\rho_{0}+AT^{2}$.
There exists a large anisotropy in the resistivity of ${\rm CeIrIn_{5}}$.
This anisotropy is closely  related to the quasi-two dimensional
Fermi surface. 
The band calculations were performed for ${\rm 
CeIrIn_{5}}$~\cite{rf:FSCeIr}
 and  ${\rm CeCoIn_{5}}$
~\cite{rf:FSCeCo}.
The Fermi surfaces calculated for ${\rm CeIrIn_{5}}$ and ${\rm
CeCoIn_{5}}$, respectively are almost same.
The topology of the Fermi surface 
is explained by the 4$f$-itinerant band model.
The calculated Fermi surfaces consist
of nearly cylindrical Fermi surfaces and small ellipsoidal ones.

\subsection{Model Hamiltonian for ${\rm CeIr}_{x}{\rm Co}_{1-x}{\rm In}_{5}$}
We calculate here $T_{{\rm c}}$ 
of ${\rm CeIr}_{x}{\rm Co}_{1-x}{\rm In}_{5}$.
According to the discussion in $\S 2$,
we start from the quasi-particle state and then calculate 
the momentum dependent 
interaction between quasi-particles by the perturbation theory and 
derive the superconductivity of 
${\rm CeIr}_{x}{\rm Co}_{1-x}{\rm In}_{5}$.
Taking the importance of the most heavy quasi-two dimensional
Fermi surface with $f$-character into account in realizing the
superconductivity in ${\rm CeIr}_{x}{\rm Co}_{1-x}{\rm In}_{5}$, 
we represent the band as an effective 
two-dimensional Hubbard model on the tetragonal lattice.
In this section, we 
call quasi-particle ``electron''and call the renormalized 
on-site repulsion between quasi-particles $\tilde{\Gamma}_{{\rm loc}}$ 
`` Coulomb repulsion $U$ ''.
We consider only the nearest neighbor (effective) hopping integrals $t$.
We rescale length, energy, temperature, time by
$a, t, \frac{t}{k_{\rm B}}, \frac{\hbar}{t}$ respectively
(where $a, k_{\rm B}, \hbar$ are the  lattice constant of the tetragonal
basal plane, Boltzmann constant, Planck constant divided by $2\pi$
respectively), we write our model Hamiltonian as follows:
\begin{eqnarray}
H&=&H_{0}+H_{1},\\
H_{0}&=&\sum_{{\bf k},\sigma}\left(\epsilon
({\bf  k})-\mu
\right)
a_{{\bf k}\sigma}^{\dagger}a_{{\bf k}\sigma},
\end{eqnarray}
\begin{equation}
\epsilon({\bf  k})=2(\cos k_{x}+\cos k_{y}),
\end{equation}
\begin{equation}
H_{1}=\frac{U}{2N}
\sum_{\sigma\neq\sigma^{\prime}}
\sum_{{\bf k}_{1}{\bf k}_{2}{\bf k}_{3}
{\bf k}_{4}}
\delta_{{\bf k}_{1}+{\bf k}_{2},
{\bf k}_{3}+{\bf k}_{4}}
a_{{\bf k}_{1}\sigma}^{\dagger}
a_{{\bf k}_{2}\sigma^{\prime}}^{\dagger}
a_{{\bf k}_{3}\sigma^{\prime}}a_{{\bf k}_{4}\sigma},
\end{equation}
where $a_{{\bf k}\sigma}^{\dagger}(a_{{\bf k}\sigma})$
is the creation(annihilation) operator for the electron with
momentum ${\bf  k}$ and spin index $\sigma$; 
$t$ and $\mu$ are 
the hopping integral and the chemical potential, 
respectively. The sum over ${\bf  k}$ indicates taking summation over a
primitive cell of the inverse lattice.
The Fermi surfaces calculated for ${\rm CeIrIn_{5}}$ and ${\rm
CeCoIn_{5}}$, respectively are almost same.
Therefore we consider that the electron number $n$ per one spin site
and the Coulomb repulsion 
$U$ are determined corresponding to the composition $x$.
\subsubsection{model parameters}
Our model parameters are the Coulomb repulsion 
$U$ and the electron number $n$ per one spin site.
The bandwidth $\Delta\epsilon$ of our model is $8$.
According to the band calculation and 
the de Haas-van Alphen  experiment~\cite{rf:FSCeIr,rf:FSCeCo}, 
the parameter region of our model which reproduces the Fermi sheet is
given by $0.345\le n\le 0.385$. 
We consider that $n=0.345$ and $n=0.385$ are the best
fitting values of parameters which well reproduce 
the Fermi sheet of ${\rm CeIrIn_{5}}$ and ${\rm CeCoIn_{5}}$, respectively.
The superconducting transition temperatures $\tilde{T}_{{\rm c}}$ 
of ${\rm CeIrIn_{5}}$ 
and ${\rm CeCoIn_{5}}$ determined 
by experiments are about 0.4 K and 2.3 K, respectively. 
To evaluate the rescaled $T_{{\rm c}}(=k_{{\rm B}}\tilde{T}_{{\rm c}}/t)$
, we have to estimate the 
renormalized bandwidth $\Delta\epsilon(=8t)$ 
because we have started from the 
renormalized Fermi liquid.
We estimate $\Delta\epsilon$ by using following equation;
$\Delta\epsilon\simeq\frac{m_{{\rm band}}}{m_{{\rm dHvA}}}
\Delta\epsilon_{{\rm band}}$, 
where $m_{{\rm band}}, m_{{\rm dHvA}}, \Delta\epsilon_{{\rm band}}$ 
are 
the effective mass calculated by the band 
calculation~\cite{rf:FSCeIr,rf:FSCeCo},
the cyclotron effective mass obtained the de Haas-van Alphen by
 experiment~\cite{rf:FSCeIr,rf:FSCeCo}, the bandwidth calculated by the 
band calculation~\cite{rf:FSCeIr,rf:FSCeCo}, respectively.
Then the rescaled superconducting transition
temperatures of  ${\rm CeIrIn_{5}}$ and ${\rm CeCoIn_{5}}$ are,
respectively, $T_{\rm c( CeIrIn_{5})}\simeq 3.2\times 10^{-3}$ 
and $T_{\rm c( CeCoIn_{5})}\simeq 3.3\times 10^{-2}$ 
when we set $0.1$ Ry as the value of 
$\Delta\epsilon_{{\rm band}}^{{\rm CeIrIn_{5}}}(\simeq
\Delta\epsilon_{{\rm band}}^{{\rm CeCoIn_{5}}})$ according to 
the band calculation~\cite{rf:FSCeIr}.

\subsection{Green's functions}
\subsubsection{Bare Green's functions}
The bare Green's function is the following,
\begin{equation}
G^{(0)}(k)
=\frac{1}{{\rm i}\epsilon_{n}-\epsilon({\bf  k})+\mu},
\end{equation}
where $\epsilon_{n}=(2n+1)\pi T(n:{\rm integer})$ is the fermion-Matsubara
frequency and short notation
$k=({\bf  k},\epsilon_{n})$ is adopted.
We consider the Hartree term is included in the chemical potential.
\subsubsection{Dressed normal Green's functions}
\begin{figure}
\includegraphics[width=0.6 \linewidth]{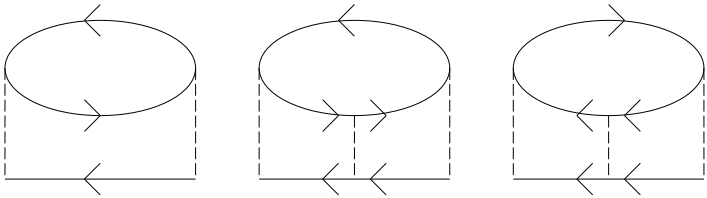}
\caption{The Feynman diagrams of the normal self-energy up to third
 order. The solid and dashed lines correspond to the bare Green's function 
 and the interaction, respectively.}
\label{fig:NSE}
\end{figure}

Next we consider the dressed normal Green's function.
When we consider the situation near the superconducting 
transition temperature,
the Dyson-Gorkov's equation can be linearized. Therefore 
the dressed normal Green's function is obtained from the bare Green's
function with only the  normal self-energy correction.
We expand the normal self-energy  up to third order with
respect to $U$;the diagrams are shown in Fig.~\ref{fig:NSE}.

Then we obtain the normal self-energy as follows,
\begin{equation}
\Sigma_n(k)=
\frac{T}{N}\sum_{k^{\prime}}
[U^{2}\chi_0(k- k^{\prime})
+U^{3}\chi_0^{2}(k- k^{\prime})
+U^{3}\phi_0^{2}(k+k^{\prime})
]G_0(k^{\prime}),
\end{equation}

where $\chi_0(\cdots)$ and $\phi_0(\cdots)$ are given respectively as
\begin{equation}
\chi_0({\bf  q},\omega_m)=-\frac{T}{N}\sum_{{\bf k},n}
G_0({\bf  k},\epsilon_n)G_0({\bf  q}+{\bf  k},\omega_m+\epsilon_n),
\end{equation}
\begin{equation}
\phi_0({\bf  q},\omega_m)=-\frac{T}{N}\sum_{{\bf k},n}
G_0({\bf  k},\epsilon_n)G_0({\bf  q}-{\bf  k},\omega_m-\epsilon_n).
\end{equation}
Here $\omega_{m}=2m\pi T\nonumber(m:{\rm integer})$
 is the boson-Matsubara frequency. 
The quantity $\chi_{0}({\bf  q},\omega_{m})$ has
the physical meaning of the bare susceptibility and expresses spin
fluctuations in the system. More over, the bare susceptibility plays an
important role in the calculation of $T_{\rm c}$, that is, it determines the 
magnitude and the spatial and temporal variation of the effective
interaction between electrons,
through the higher 
order terms in  $U$. 
(See the first term of right hand side of the equation ~\ref{Eli}.)

Note that the Hartree term has been already included in the chemical potential
and the constant terms which have not been included in the Hartree term  are
included in the chemical potential shift when we fix the particle
number.

Then the dressed normal Green's function is 
\begin{equation}
G(k)=\frac{1}{{\rm i}\epsilon_n-(\epsilon({\bf  k})-\mu-\delta\mu+\Sigma_n(k))},
\end{equation}
where $\delta\mu$ is determined so that the following equation is satisfied.
\begin{equation}
n=\frac{T}{N}\sum_{k}G(k)=\frac{T}{N}\sum_{k}G_0(k).
\end{equation}
We expand the above equation up to the third order
of the interaction with regard to $\delta\mu-\Sigma_n(k)$, we obtained  $\delta\mu$ as 
\begin{equation}
\delta\mu=-\frac{\frac{T}{N}\sum_{k}G_0^{2}(k)\Sigma_n(k)}{\chi_0({\bf 0},0)}.
\end{equation}

\subsubsection{Anomalous self-energy and effective interaction}
\begin{figure}
\includegraphics[width=0.6 \linewidth]{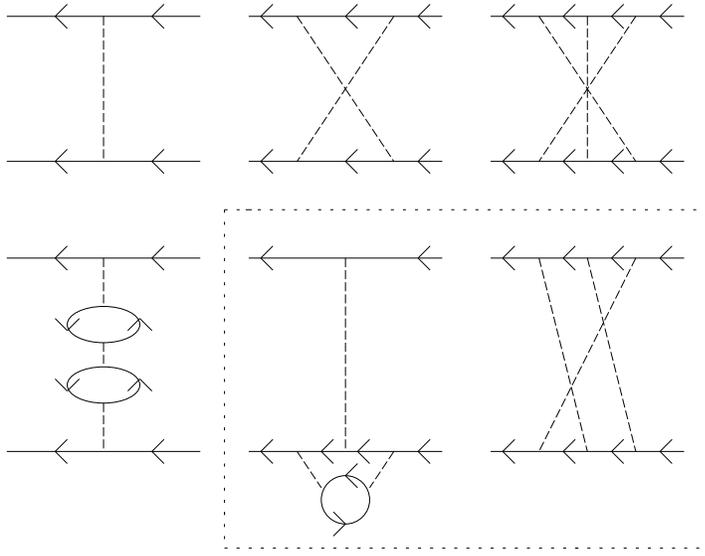}
\caption{The Feynman diagrams of the effective interaction up to third
 order. The solid and dashed lines correspond to the bare Green's function 
 and the interaction, respectively. The diagrams enclosed by the dashed
 line are vertex corrections. The other diagrams are included in RPA.}
\label{fig:EI}
\end{figure}

When we consider the situation
near the superconducting transition temperature, the anomalous self-energy $\Sigma_{a}(k)$ is 
represented by the anomalous Green's function $F(k)$ and the (normal)effective
interaction. We expand the effective interaction up to third order with
respect to $U$ as shown in Fig.~\ref{fig:EI}.

Then we obtain the anomalous self-energy as follows;
\begin{eqnarray}
\Sigma_a(k)&=&
-\frac{T}{N}\sum_{k^{\prime}}[
U+U^{2}\chi_0(k+k^{\prime})+2U^{3}\chi_0^{2}(k+k^{\prime})
]F(k^{\prime})\nonumber\\
& &\mbox{}-U^{3}\frac{T^{2}}{N^{2}}\sum_{k^{\prime}k^{\prime\prime}}
G_0(k^{\prime})[\chi_0(k+k^{\prime})-\phi_0(k+k^{\prime})]
G_0(k+k^{\prime}-k^{\prime\prime})F(k^{\prime\prime})
\nonumber\\
& &\mbox{}
-U^{3}\frac{T^{2}}{N^{2}}
\sum_{k^{\prime}k^{\prime\prime}}
G_0(k^{\prime})[\chi_0(-k+k^{\prime})-\phi_0(-k+k^{\prime})]G_0(-k+k^{\prime}-k^{\prime
\prime})F(k^{\prime\prime}).
\end{eqnarray}

\subsection{$\acute{E}$liashberg's equation}
From the linearized Dyson-Gorkov equation, we obtain the anomalous Green's
 function as follows;
\begin{equation}
F(k)=|G(k)|^{2}\Sigma_a(k).
\end{equation}
Then the ${\rm\acute{E}liashberg}$'s equation is given by

\begin{eqnarray}\label{Eli}
\Sigma_a(k)&=&-\frac{T}{N}\sum_{k^{\prime}}[
U+U^{2}\chi_0(k+k^{\prime})+2U^{3}\chi_0^{2}(k+k^{\prime})
]|G(k^{\prime})|^{2}\Sigma_a(k^{\prime})\nonumber\\
& &\mbox{} -U^{3}\frac{T^{2}}{N^{2}}\sum_{k^{\prime}k^{\prime\prime}}
G_0(k^{\prime})[\chi_0(k+k^{\prime})-\phi_0(k+k^{\prime})]G_0(k+k^{\prime}-k^{\prime\prime})|
G(k^{\prime\prime})|^{2}
\Sigma_a(k^{\prime\prime})
\nonumber\\
& &\mbox{}
-U^{3}\frac{T^{2}}{N^{2}}
\sum_{k^{\prime}k^{\prime\prime}}
G_0(k^{\prime})[\chi_0(-k+k^{\prime})-\phi_0(-k+k^{\prime})]G_0(-k+k^{\prime}-k^{\prime\prime})|G(k^{\prime\prime}
)|^{2}\Sigma_a(k^{\prime\prime}).\nonumber\\
\end{eqnarray}
This equation is corresponding to the Eq.( \ref{qp-kuriEli} ).
We consider that the system is superconducting state 
when the eigen-value of this equation is 1.
\subsection{Calculation Results }
\subsubsection{Details of the numerical calculation}
To solve the ${\rm\acute{E}}$liahberg's equation by using the
power method algorithm, we have to calculate the 
summation over the momentum and the frequency space. Since all summations
 are in the convolution forms, we can carry out them by using the
 algorithm of the Fast Fourier Transformation.
For the frequency, irrespective of the temperature, 
we have 1024 Matsubara
frequencies. Therefore we calculate throughout 
in the temperature region $T\ge T_{\rm lim}$
, where $T_{\rm lim}$ is the lower limit temperature
 for reliable numerical calculation,
which is estimated about $2.0\times 10^{-3}
(>\Delta\epsilon/(2\pi\times 1024)\simeq 1.2\times 10^{-3})$;
 we divide a primitive cell into 128$\times$128 meshes.

We have carried out analytically continuing procedure 
by using  Pad${\rm\acute{e}}$ approximation.
\subsubsection{Dependence of $T_{\rm c}$ on $U,n$ and vertex correction terms}
To solve the ${\rm \acute{E}}$liashberg's
equation, we set the initial gap function
($d_{x^{2}-y^{2}}$-symmetry) as follows.
\begin{equation}
\Sigma_{a}(k)\propto \cos k_{x}-\cos k_{y}.
\end{equation}

\begin{figure}
\includegraphics[width=1 \linewidth]{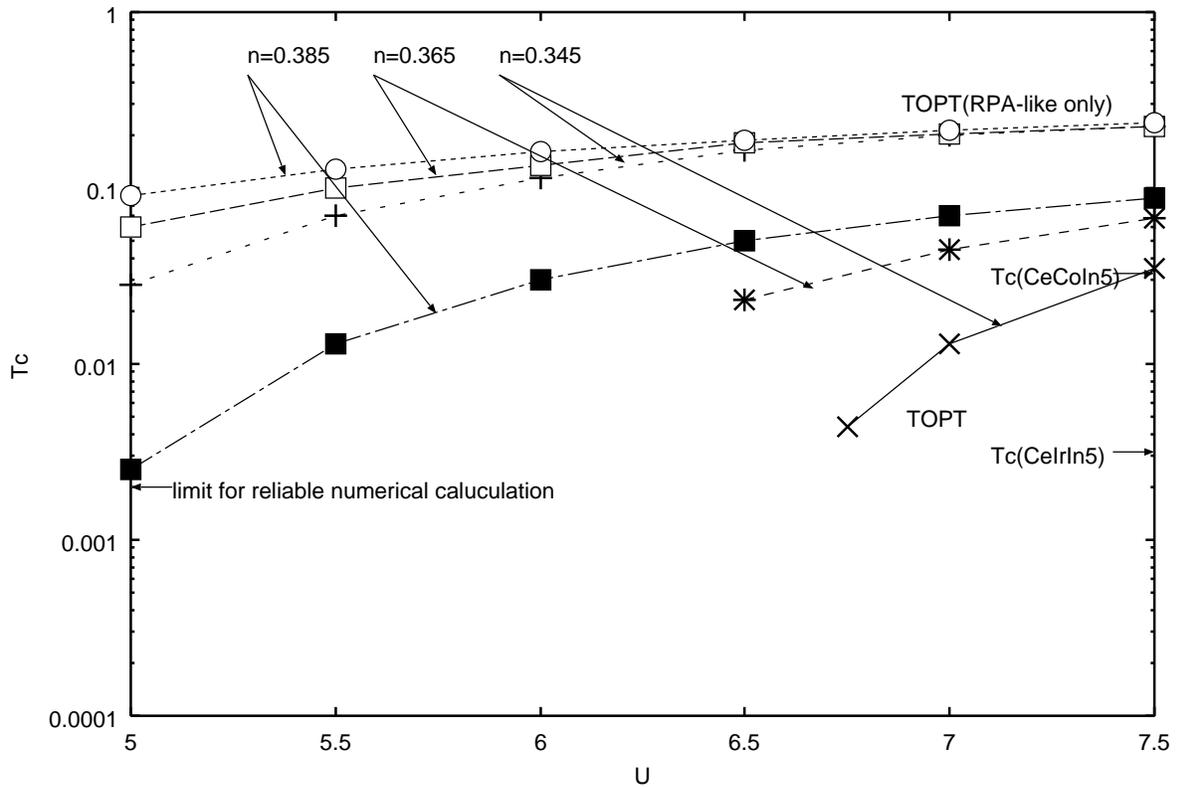}
\caption{The calculated $T_{\rm c}$ as $U$ is varied and 
for various  values of $n$ as shown in the figure. 
TOPT and TOPT(RPA-like only) in this figure mean
 that full diagrams and only RPA-like diagrams of 
anomalous self-energies up to third order are included, respectively.}
\label{fig:Tc}
\end{figure}

\begin{figure}
\includegraphics[width=1 \linewidth]{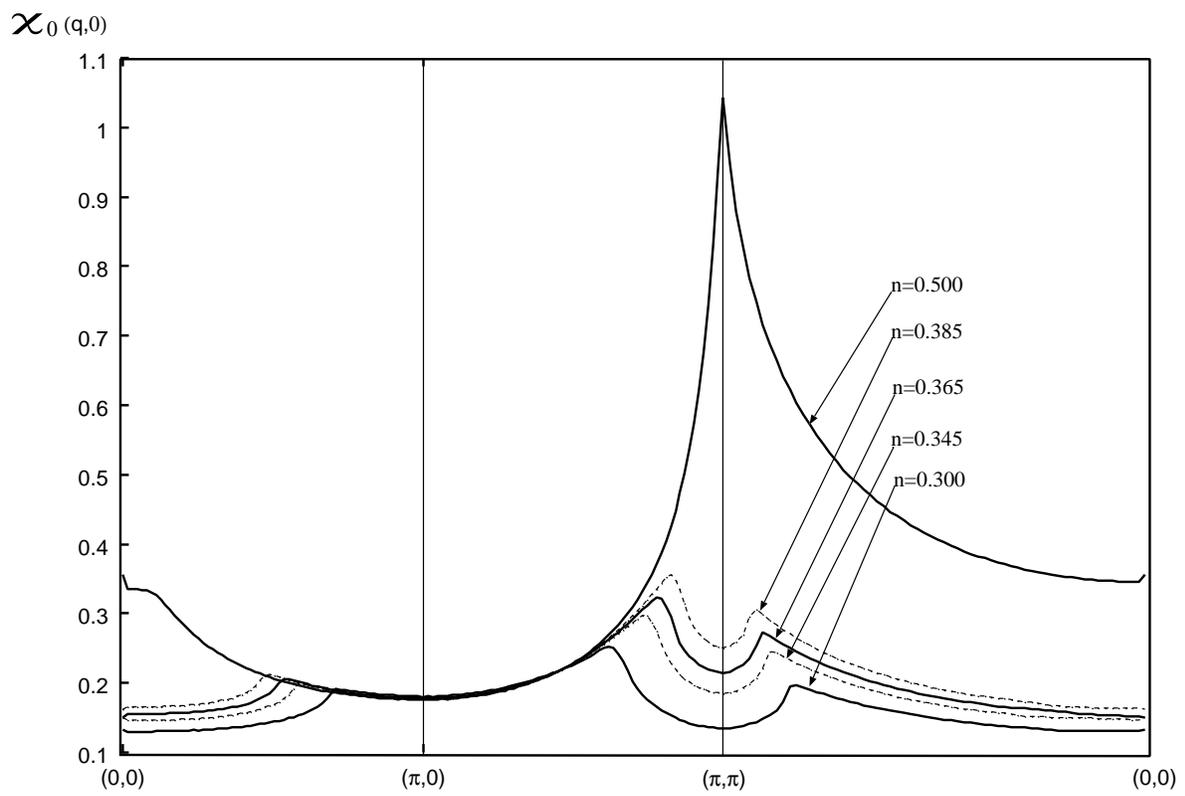}
\caption{The momentum dependence of the static 
bare susceptibility for various $n$.}
\label{fig:X0}
\end{figure}

\begin{figure}
\includegraphics[width=1 \linewidth]{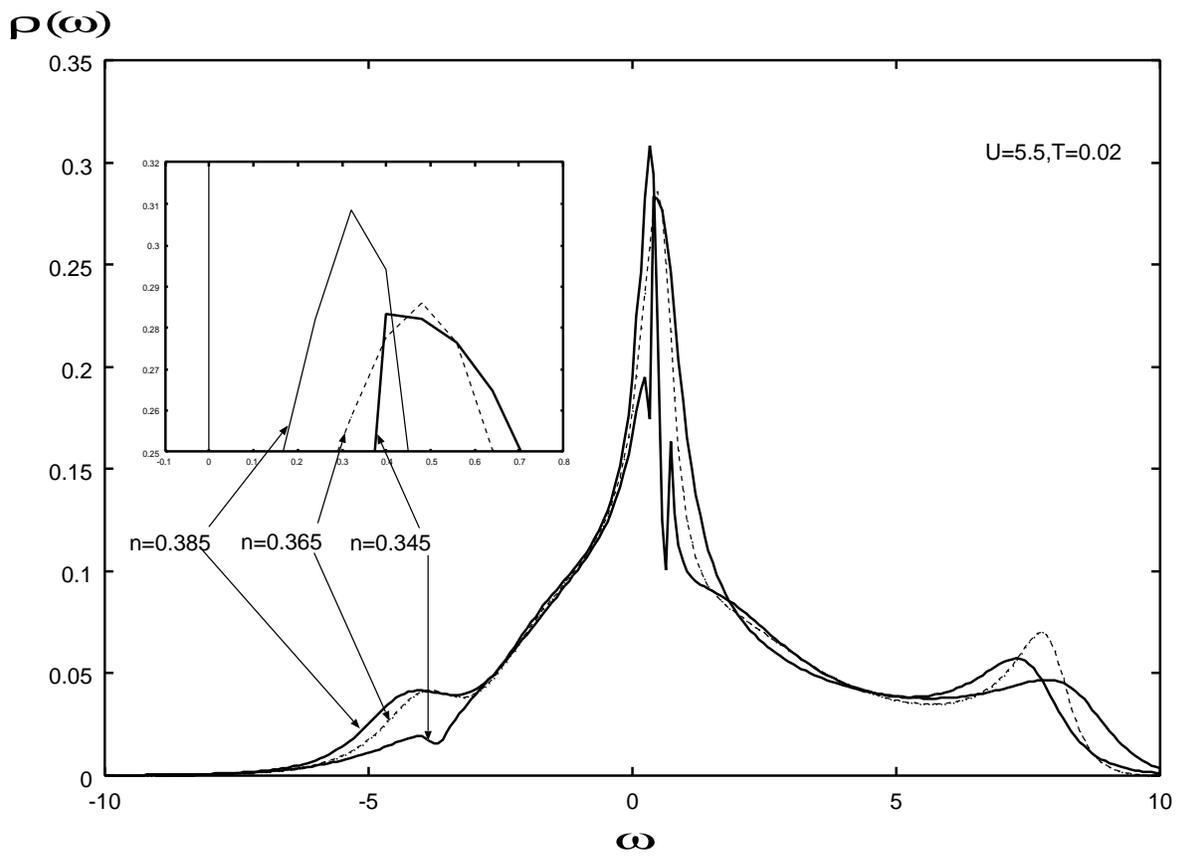}
\caption{The density of states as $n$ is varied, at 
$U=5.5$ and $T=0.02$. The inset shows the details near  the Fermi level.}
\label{fig:DOS}
\end{figure}

The calculated gap functions show the node at $k_{x}=k_{y}$ and
$k_{x}=-k_{y}$ and changes the sign across the node in all approximations 
and for all parameters. 
The symmetry of Cooper pair is $d_{x^{2}-y^{2}}$.
The dependence of $T_{\rm c}$ on $U, n$ are shown in Fig.~\ref{fig:Tc}.
To examine how the vertex corrections influence  $T_{\rm c}$,
we also calculate $T_{\rm c}$ by including only the RPA-like
 diagrams of anomalous self-energies up to third order, 
in other words, without the vertex corrections.
We compare obtained $T_{\rm c}$(RPA-like only) 
with $T_{\rm c}$ calculated by 
including full diagrams of anomalous self-energies up to third order, in
 Fig.~\ref{fig:Tc}.
From this figure, we can point out the following 
facts. 
For large $U$ higher $T_{\rm c}$ are 
obtained commonly for all cases.
$T_{\rm c}$ calculated by including only the RPA-like
 diagrams of anomalous self-energies up to third order is higher than 
$T_{\rm c}$ calculated by including full diagrams of them. 
These results show that the main origin of the d-wave superconductivity
is the momentum and frequency dependence of  
spin fluctuations given by the RPA-like terms.
The vertex corrections reduce $T_{\rm c}$ by one order of magnitude.
So the vertex corrections is important for obtaining reasonable $T_{\rm
 c}$ and for differentiating $n$-dependence of $T_{{\rm c}}$. 
When we fix the Coulomb repulsion $U$ and increase 
$n(0.345 \sim 0.385)$, the system get close to the
half-filling state ($n=0.5$). In this case, 
the momentum dependence of $\chi_{0}({\bf  q},0)$ is slightly
enhanced (see Fig.~\ref{fig:X0} shown later) 
and at the same time Fermi level gets close
to the van Hove singularity (see Fig.~\ref{fig:DOS} shown later), 
then higher $T_{\rm c}$ are obtained.
If we assume $U_{{\rm CeIrIn_{5}}}\simeq U_{{\rm CeCoIn_{5}}}$, 
$T_{{\rm c}}$ of ${\rm CeIrIn_{5}}$ with $n=0.345$ is lower than that of 
${\rm CeCoIn_{5}}$ with $n=0.385$ and this tendency is in good agreement
with experimental results.
\subsubsection{Behavior of $\chi_{0}({\bf  q},0)$}
The calculated results of the static bare susceptibility are shown
in Fig.~\ref{fig:X0} for various values of $n$.
From this figure, we point out the following facts.
In the case of half-filling state, $\chi_{0}({\bf  q},0)$ has commensurate
peak at $(\pi,\pi)$.
When we increase $n(0.3 \sim 0.5)$, 
the system get close to the half-filling state and the peak 
and the momentum dependence of $\chi_{0}({\bf  q},0)$ are enhanced.
In the case of $n=0.345 \sim 0.385$, 
$\chi_{0}({\bf  q},0)$ has the incommensurate peak around $(\pi,\pi)$.
This peak is not prominent but $\chi_{0}({\bf  q},0)$ has
sufficiently strong  momentum dependence.
These results mentioned above indicates 
the $d_{x^{2}-y^{2}}$ symmetry of the gap function.
\subsubsection{Density of states}
The density of states(DOS) is given by 
\begin{equation}
\rho(\omega)=-\frac{1}{N\pi}\sum_{{\bf k}}{\rm Im}G^{R}({\bf  k},\omega),
\end{equation}
where
\begin{displaymath}
G^{R}({\bf  k},\omega)=\left.G({\bf  k},\epsilon_{n})\right|_{{\rm i}\epsilon_{n}\rightarrow \omega+{\rm i}\eta}.
\end{displaymath}
We show
the $n$-dependence of DOS in Fig.~\ref{fig:DOS}.
From inset in this figure,
we can see that the position of the van Hove singularity shifts
upward from the Fermi
level with decreasing $n$.
This departure of the van Hove singularity from the Fermi level reduces
the superconducting transition temperature $T_{{\rm c}}$.

\subsubsection{Self-energy}
The self-energy is given by
\begin{displaymath}
\Sigma_{n}^{R}({\bf  k},\omega)=\left.\Sigma_{n}({\bf  k},\epsilon_{n})\right|_{{\rm i}\epsilon
_{n}\rightarrow\omega+{\rm i}\eta}.
\end{displaymath}

\begin{figure}
\includegraphics[width=1 \linewidth]{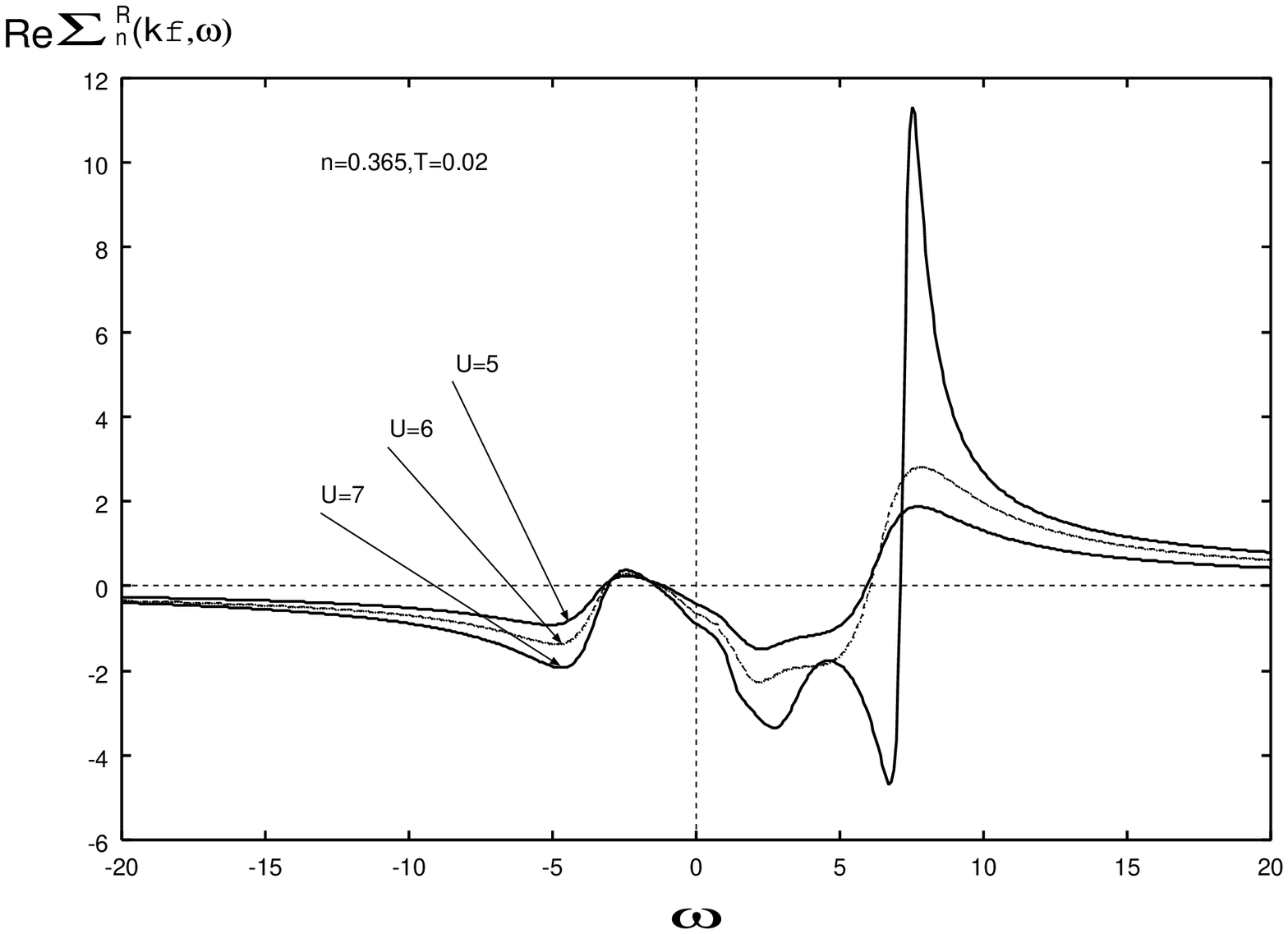}
\caption{The real part of the normal self-energy at the Fermi momentum,
 at $n=0.365$, $T=0.02$ and for 
various values of $U$ as shown in the figure.}
\label{fig:ReSE}
\end{figure}

\begin{figure}
\includegraphics[width=1 \linewidth]{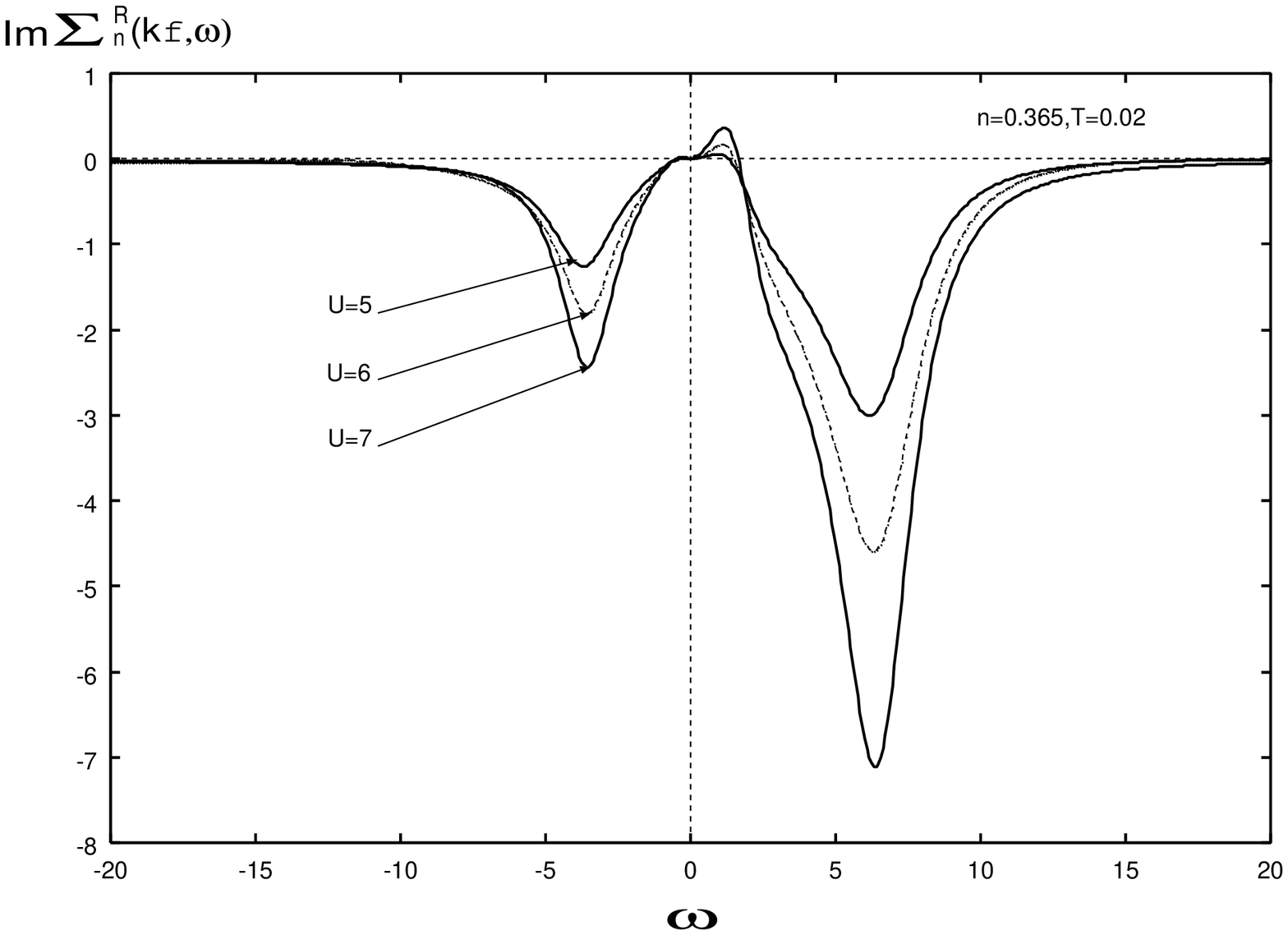}
\caption{The imaginary part of the normal self-energy at the Fermi momentum,
 at $n=0.365$, $T=0.02$ and 
for various values of $U$ as shown in the figure.}
\label{fig:ImSE}
\end{figure}

The real part and the imaginary part of the self-energy at Fermi
momentum are shown in 
Fig.~\ref{fig:ReSE} and Fig.~\ref{fig:ImSE} respectively.
The $\omega$-dependence of both parts near $\omega=0$ are respectively 
given by 
${\rm Re}\Sigma_{n}^{R}({\bf  k}_{f},\omega)\propto -\omega$ and 
${\rm Im}\Sigma_{n}^{R}({\bf  k}_{f},\omega)\propto -\omega^{2}$.
This behavior is the same as that for the usual Fermi liquid.
As $U$ increases, the slope of ${\rm Re}\Sigma_{n}^{R}({\bf 
k}_{f},\omega)$ at $\omega=0$ becomes steeper and the coefficient of the
$\omega^{2}$-term in ${\rm Im}\Sigma_{n}^{R}({\bf  k}_{f},\omega)$ become
larger. This indicates that the mass and the damping rate of the
quasi-particle become larger as $U$ increases.
These results are the typical Fermi liquid ones.
\section{Summary, Discussion and Conclusion}
In this paper, we have reformulated 
the superconducting transition on the quasi-particle
description and discussed of the superconductivity 
of ${\rm CeIr}_{x}{\rm Co}_{1-x}{\rm In}_{5}$ 
on the basis of such a renormalized formula.
By the low order perturbation expansion it is difficult to obtain actual 
mass enhancement factor. Here we have started from heavy fermions
possessing a large effective mass. To obtain the anisotropic
superconductivity we have taken account of the momentum dependence
of interaction between quasi-particles. By this procedure we can make an 
argument consistent with both heavy electron 
mass and anisotropic superconductivity.
Taking the importance of the most heavy quasi-two dimensional
Fermi surface into account in realizing the
superconductivity in ${\rm CeIr}_{x}{\rm Co}_{1-x}{\rm In}_{5}$, 
we have represented these bands with an effective single band 
two-dimensional Hubbard model and calculated
$T_{\rm c}$ on the basis of the third order perturbation theory.
We have pointed out that 
the main origin of their superconductivities can be ascribed to 
the momentum and frequency dependence of
the spin fluctuations, although
the vertex correction terms are important for 
reducing $T_{\rm c}$ then  obtaining reasonable transition temperatures 
and differentiating $n$-dependence of $T_{{\rm c}}$.

Almost the same results have been obtained in the calculation of
$T_{{\rm c}}$ for
high-$T_{\rm c}$ cuprates performed by Hotta
~\cite{rf:HoTOPT}
and for organic superconductors performed by Jujo {\it et al.}
~\cite{rf:JuTOPT}, 
based on the third order perturbation theory.
For high-$T_{\rm c}$ cuprates, Hotta calculated $T_{{\rm c}}$ 
on the basis of $d$-$p$ model and succeeded in 
reproducing experimentally obtained dependence of $T_{{\rm c}}$
on carrier number for overdoped cuprate within 
the third order perturbation theory. 
In spin-triplet
superconductor ${\rm Sr_{2}RuO_{4}}$, 
Nomura and Yamada have recognized that 
the momentum and frequency dependence of 
the effective interaction between electrons
which is not included in the interaction mediated by the spin fluctuation 
is essential for realizing the spin triplet pairing,
by treating the electron correlation
within the third order perturbation theory, on the basis of the
two-dimensional  single-band
Hubbard model.~\cite{rf:NoTOPT}.
According to their theory, third order vertex correction terms, which are
not direct contribution from spin fluctuations, are important for the
superconductivity in ${\rm Sr_{2}RuO_{4}}$.
They also searched finite values of $T_{{\rm c}}$ within 
the second order perturbation theory (the second order term is the 
only term which is included in the interaction mediated by 
the spin fluctuation within 
the third order perturbation theory in the case of
spin-triplet superconductivity of the single-band Hubbard model) 
and the random phase approximation, 
but could not find any finite value of $T_{{\rm c}}$ within the
precision of their numerical calculations and for 
appropriate values of parameters.
Recently Kondo has examined the superconductivity of the two-dimensional 
Hubbard model in the small $U$ limit within the second perturbation
theory. One of his conclusion is also 
that the spin-triplet superconducting state is 
not stable in the small $U$ limit~\cite{rf:Kondo}.

These facts described above suggest that the calculations of $T_{{\rm c}}$
which include only the spin fluctuation terms 
are questionable and should be carefully 
performed with other terms.
In our calculation, 
the effects of other terms are taken into account by 
third order vertex correction terms which are
not included in the interaction mediated by the 
spin fluctuation.
We assume that the effect of the electron correlation is almost 
taken into account by the third order perturbation theory.
To confirm this assumption, we have to do with the forth order perturbation theory.
The forth order perturbation theory is a remaining future problem.
We point out that in
calculating the superconducting transition temperature $T_{{\rm c}}$ 
of typical heavy fermion superconductors, it is reasonable to treat the
electron correlation 
by perturbation theory based on the Fermi liquid theory 
instead of taking only the effect of 
the specific diagrams which are favorable to the spin fluctuation.

The electronic structure of pressured ${\rm CeRhIn_{5}}(P>1.6{\rm GPa})$ 
is considered to be same structure of ${\rm CeCoIn_{5}}$. So we
consider that the same mechanism can be applied to the mechanism of
the superconductivity of pressured ${\rm CeRhIn_{5}}(P>1.6{\rm GPa})$. 

In conclusion, we have reformulated the \'Eliashberg's equation for the 
superconducting transition within the quasi-particle
description  which takes into account of the heavy electron mass.
We also presented a microscopic mechanism for the d-wave
superconductivity in ${\rm CeIr}_{x}{\rm Co}_{1-x}{\rm In}_{5}$ and 
attempted to explain the $x$-dependence of $T_{{\rm c}}$, by
using such a renormalized formula and the perturbation theory.
\section*{Acknowledgments}
Numerical computation in this work was carried out at the Yukawa
 Institute Computer Facility.

\newpage

\end{document}